\documentclass[ preprint,
superscriptaddress,amsmath,amssymb,
 aps]{revtex4-2}

\usepackage[T1]{fontenc}
\usepackage{amsmath}
\usepackage{amsfonts}
\usepackage{amssymb}
\usepackage{graphicx}

\usepackage{comment}
\usepackage{float}
\usepackage{subcaption}
\usepackage{graphicx}
\usepackage{comment}
\usepackage{color}

\begin{document}

\title{Rotational dynamics of bound pairs of bacteria-induced membrane tubes}

\author{Makarand Diwe}
\affiliation{Raman Research Institute, C V Raman Avenue, Bengaluru, 560080, India}
\author{P B Sunil Kumar}
\affiliation{Department of Physics, Indian Institute of Technology, Madras, Chennai 600036}
\author{Pramod Pullarkat}
\affiliation{Raman Research Institute, C V Raman Avenue, Bengaluru, 560080, India}

\def\a{0.6}
\def\b{0.25}
\def\c{0.38}
\def\d{0.25}

\begin{abstract}

We present experiments demonstrating tube formation in giant unilamellar vesicles that are suspended in a bath of swimming \textit{E. coli} bacteria. We chose the lipids such that the bacteria have no adhereing interaction with the membrane. The tubes are generated by the pushing force exerted by the bacteria on the membrane of the vesicles. Once generated, the bacteria are confined within the tubes, resulting in long-lived tubes that protrude into the vesicle. We show that such tubes interact to form stable bound pairs that orbit each other. We speculate that the tubes are maintained by the persistent pushing force generated by the bacterium, and the rotating pairs are stabilized by a combination of curvature mediated interaction and vorticity generated in the membrane by the rotation of the flagella. 
    
\end{abstract}	

\maketitle

\section*{Introduction}
Several recent studies have shown that active particles exhibit diverse interactions with soft boundaries. For example, active colloidal particles show persistent orbital motion around  giant unilamellar vesicles (GUVs) \cite{sharma2021active, vincent2025curvature}. When active particles are confined within a vesicle, they also induce significant shape deformations such as tethers and dendritic structures  \cite{vutukuri2020active, takatori2020active}. 
Active particles interacting with a GUV can transfer both forces and torques, leading to GUV translation and rotation \cite{sharma2021active}. When encapsulated within a GUV, motile bacteria can deform the membrane by extruding tubes that, by coupling with their flagella, propel the entire vesicle \cite{takatori2020active, le2022encapsulated}. Generally, the pressure exerted by active particles on curved walls is inhomogeneous and can lead to driven shape instabilities on flexible walls. However, at least in two-dimensions, an equation of state can be recovered for the average normal force \cite{nikola2016active}. 
It is also known that asymmetric boundaries can rectify particle motion, generating currents and shear stresses, while pressure variations can destabilize flexible walls, leading to phenomena such as filament bending and self-propulsion \cite{nikola2016active}\cite{le2022encapsulated}.
	
Tubular structures emerge as a unifying theme in the strategies employed by diverse intracellular bacterial pathogens to manipulate host cell processes for their benefit \cite{robbins1999listeria, dowd2020listeria, colonne2016hijacking, martinez2018tiny}. \textit{Listeria monocytogenes} exploit host exocytic machinery to generate membrane-bound protrusions that resemble tubular extensions, facilitating its spread between epithelial cells \cite{robbins1999listeria, dowd2020listeria}. Similarly, \textit{Salmonella enterica} remodels the host membraneous organs to generate interconnected tubular membranes, including double membrane Salmonella-induced filaments, which function in nutrient acquisition, immune evasion, and maintenance of the Salmonella containing vacuoles \cite{liss2015take}. As other examples, obligate intracellular bacteria like Chlamydia and Rickettsia leverage the host actin cytoskeleton to generate actin-rich protrusions or tunnels that promote cell invasion and intercellular dissemination \cite{martinez2018tiny}. These studies collectively illustrate how pathogens co-opt host membraneous structures or form tubular networks as essential tools for intracellular survival, replication, and spread, highlighting the central role of tube-like structures in microbial pathogenesis. Thus, understanding the interaction between motile entities and soft boundaries is important both from a physical as well as biological point of view. 

Our focus here is on membrane deformations induced by an active agent and the resulting interaction between such deformations. We have performed experiments where GUVs are suspended in a dilute bath of \textit{E. coli} that are genetically modified to show persistent swimming. Lipids are chosen to deliberately avoid any adhering interaction between the bacterium and the membrane.
We show that single E. coli can generate membrane tubes in floppy GUVs and become engulfed in them to form long-lived tubes. These tubes interact when they are nearby to form stable orbiting pairs. We analyse the structure of these tubes and the dynamics of orbiting pairs. We also provide some speculative arguments for the stability of bound pairs. 
	
\section*{Materials and methods}
	
\subsection*{Bacteria culture}
We use a genetically modified laboratory strain of \textit{E. coli} bacteria known as RP5232 (donated by Judith Armitage, University of Oxford) \cite{barak1999chemotactic}. This strain has a deletion of a cheY gene ($\Delta$cheY) which makes the flagella rotate only counter-clock-wise making them persistent swimmers. The bacteria are cultured using Tryptone Broth (TB) (Cat. no. M463-500G, Himedia). The preparation of bacteria starts with suspending 10 $\mu l$ of bacterial stock in $10$ $ml$ of TB. This suspension is maintained in a shaker-incubator at $37$ $^\circ C$ at $80$ $rpm$. The culture is allowed to grow until the optical density, as measured by the OD600 method, reaches to about $0.6$. The optical density is measured using a UV-Visible spectrometer (Nanodrop 2000c, Thermoscientific). The bacteria are then pelleted in a centrifuge and resuspended in a $56$ $mM$ glucose solution to be used for further experimentation.

\subsection*{GUV preparation}
We prepare Giant Unilamellar Vesicles (GUVs) using 1-palmitoyl-2-oleoyl-sn-glycero-3-phosphocholine lipids (POPC) and 0.2 mole-percent 1,1'-Dioctadecyl-3,3,3',3'-Tetramethylindotri-carbocyanine Iodide $DiLC_{18}$ dye following a standard electroformation protocol \cite{boban2021giant}. A signal of 1 $V_{rms}$ 
at 25 Hz is applied for $3$ hours for the preparation of GUVs, at the end of which a signal of $0.3$ $V_{rms}$ at $3$ Hz is applied to detach the vesicles from the ITO plates. The GUVs are prepared in a $50$ $mM$ of sucrose solution to make them heavier than the surrounding medium. The GUVs then settle down on the bottom surface of the sample chamber making observation using an inverted microscope easier. After preparation, the GUVs are added to the bacterial suspension and observed using video microscopy. The observation chamber consists of a bottom coverglass and a top quartz plate with the two plates separated by an aluminum spacer with air vents (groves in the plate) to support aerobic breathing mode of \textit{E. coli}. To prevent vesicles from adhering, the coverglass is coated with Bovine Serum Albumin.

\subsection*{Video microscopy}

Time-lapse video microscopy of GUVs in the presence of suspended bacteria is performed using an inverted microscope (Olympus IX70) equipped with a CCD camera (CoolSnap-EZ, Photometrics). Recordings are made at 7 frames-per-second either in fluorescence or phase-contrast mode using a 60$\times$ objective. 

\section*{Observations}

\begin{figure}
\centering
\includegraphics[width=0.8\linewidth]{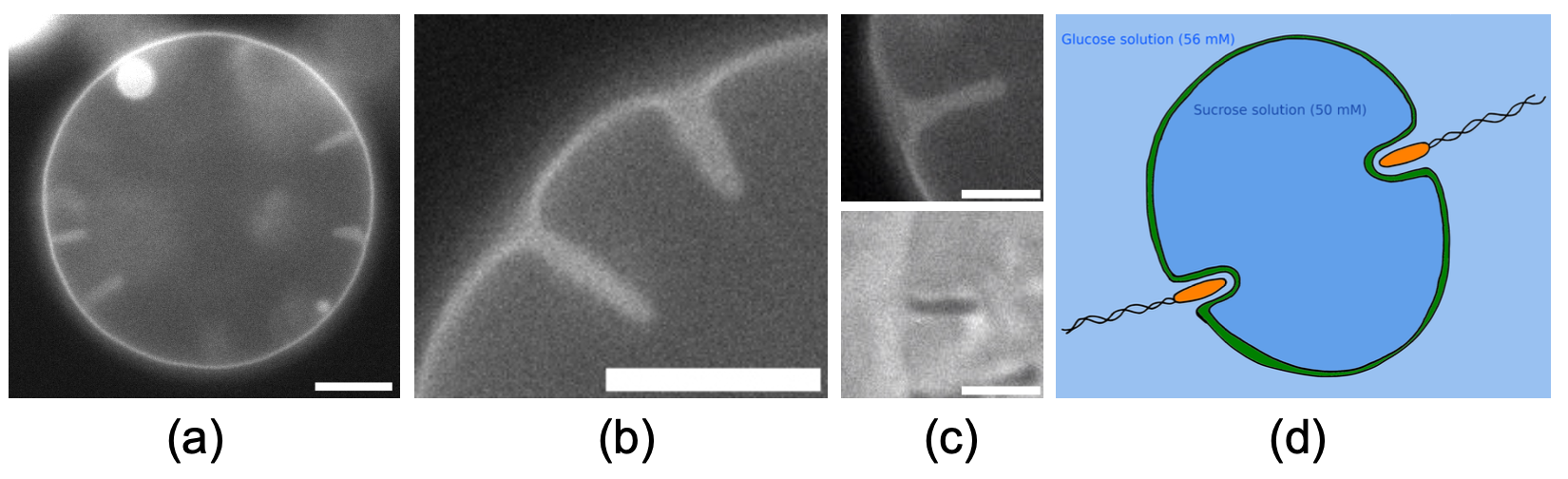}
\caption{
(a) An example of tube formation in a membrane vesicle suspended in a bath of \textit{E. coli}. (b) An enlarged view of a couple of tubes. (c) Comparison of fluorescence (above) and phase-contrast (below) images show the presence of a bacterium inside the tube. Note that the two images were taken one after the other and the tube position has shifted during this period. (d) A schematic showing the deformation of the vesicle by bacteria (not to scale).
Scale bars are 10 $\mu m$
}
\label{fig:single-tube}
\end{figure}

Giant Unilamellar Vesicles (GUVs) maintained in an external bacterial bath consisting of genetically modified \textit{E. coli} (RP5232, \cite{barak1999chemotactic}), that are persistent swimmers exhibit interesting shape dynamics. We use a neutral lipid (POPC) in order to avoid adhesive interactions of bacteria with the membrane and to investigate the vesicle deformations arising out of forces and torques exerted by the bacteria. It is known from earlier studies that E. coli adheres poorly to pure POPC membrane \cite{dai2022lipid}. The swimming forces exerted by the bacteria on the membrane surface results in the formation of membrane tubes or invaginations that protrude into the vesicle. An example of tubes formed by such interaction, imaged using fluorescence microscopy, is shown in Fig. \ref{fig:single-tube}a (also see Video-1 and Video-2 in Supplementary Material), and an enlarged view of two such tubes is shown in Fig. \ref{fig:single-tube}b. The length of the tube can vary from tube to tube. The base of the tubes can drift along the envelope of the vesicle, but the tubes always remain nearly normal to the envelop (pointed towards the GUV center). Phase contrast images show that each of these tubes engulf at least one  bacterium as shown in Fig. \ref{fig:single-tube}c. A schematic of a vesicle with bacteria containing tubes is shown in Fig. \ref{fig:single-tube}d. Usually, the number of bacteria containing tubes are higher for floppy vesicles as compared to relatively less floppy ones. This is particularly clear in cases where a given vesicle transforms from a floppy state to a tense state via extrusion of membrane, as can be seen in Video-3 in Supplementary Material. Figure \ref{fig:L&D_PDF_new} shows the distributions of lengths and diameters of tubes obtained using multiple GUV-bacteria preparations. It can be seen that the typical tube length is about 4.5 $\mu m$ and the diameter peaks at around 1.4 $\mu m$. 

\begin{figure}[h]
\centering
\includegraphics[width=0.5\linewidth]{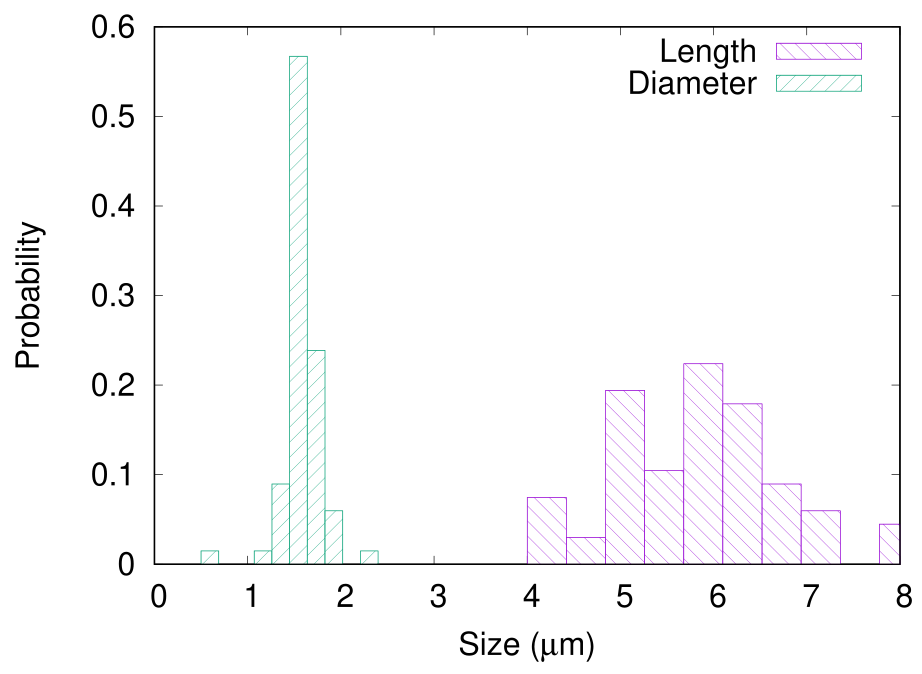}
\caption{
Probability distributions of the tube diameter and length. A total number of 67 tubes were analyzed to construct these plots. 
}
\label{fig:L&D_PDF_new}
\end{figure}

When there are extraneous objects stuck to the exterior of the vesicle, close to the base of a tube, such objects are seen to move around the tube axis. This suggest that there is circular hydrodynamic flow either within the lipid bilayer or in the external fluid, presumably generated by the rotation of the bacterial flagella. One such example is shown in Fig. \ref{fig:rotating particle} and in Video-4 in Supplementary Material. 

\begin{figure}[ht!]
\centering
\includegraphics[width=0.6\linewidth]{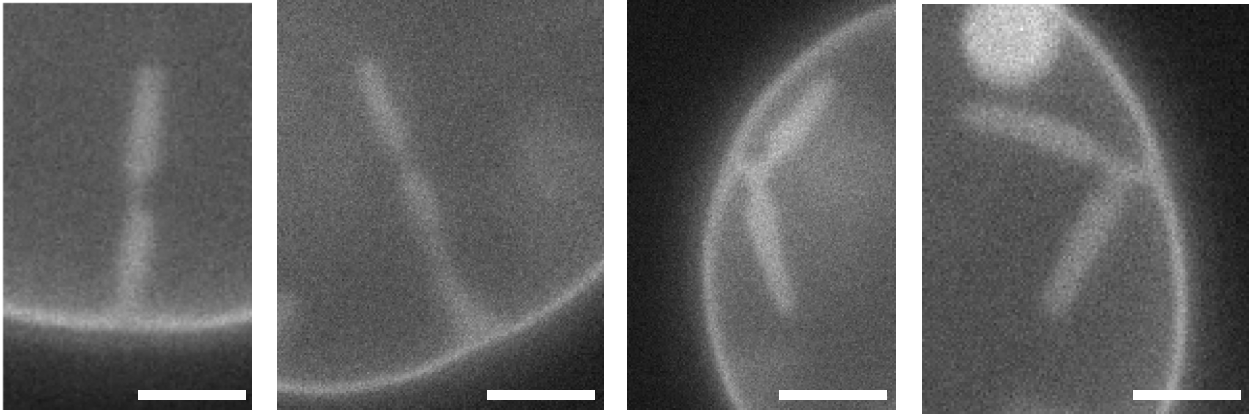}
\caption{
Example images of nearly straight tubes containing more than one bacterium (left two images) and ones with branches (right two images). Note that the tubes show narrow necks in the region between bacteria in such cases. The scale bars are 5 $\mu m$.
}
\label{fig:multiple-bacteria}
\end{figure}

In some cases, multiple bacteria enter a tubes and the tubes can get longer or even acquire branched geometries. In such cases, the tube appears slightly pinched between the two bacteria as shown in Fig. \ref{fig:multiple-bacteria}. Pinching is also seen towards the base of long tubes containing a single bacterium as in Fig. \ref{fig:single-tube}b.

\begin{figure}[h]
\centering
\includegraphics[width=0.6\linewidth]{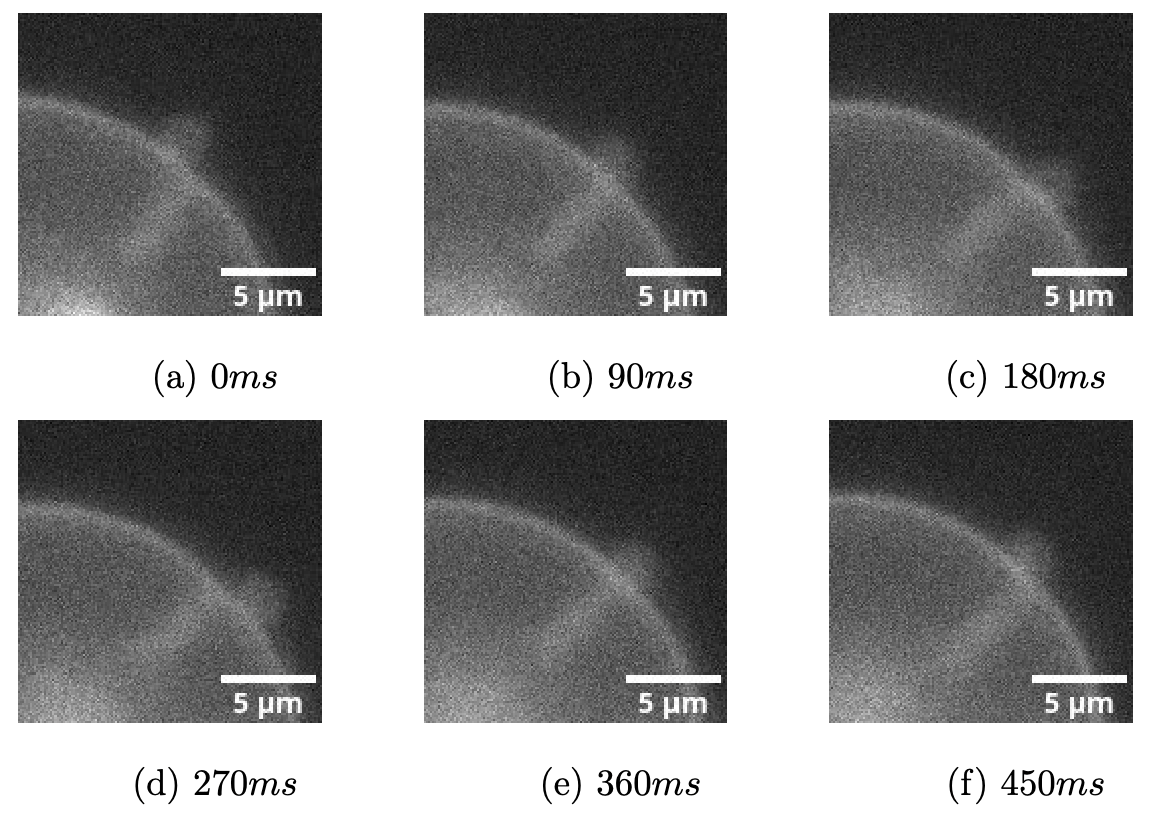}
\caption{
Time series of a vesicle bud moving around the base of a tube generated by a bacterium. This set of figures show a complete cycle of the bud around the base.
}
\label{fig:rotating particle}
\end{figure}

\begin{figure}
\centering
\includegraphics[width=1.0\linewidth]{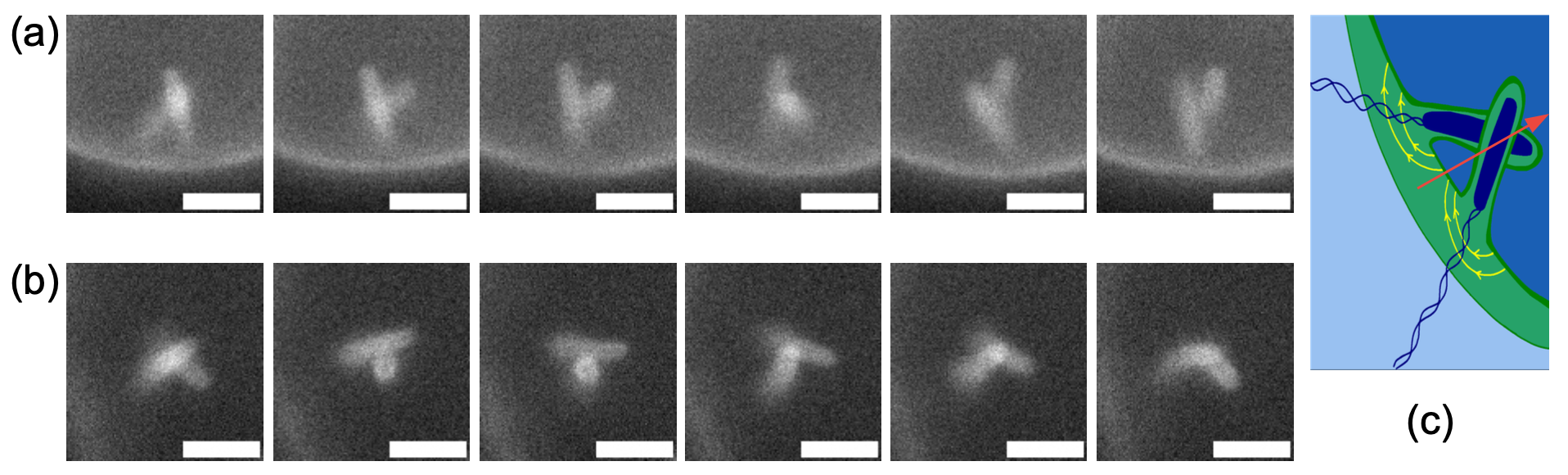}
\caption{
(a) Image sequence showing two tubes that orbit each other as seen nearly side on. The sequence shows almost half an orbit. Scale bars are 5 $\mu m$. (b) Image sequence showing a pair of orbiting tubes as seen almost along the axis of the orbit for almost half an orbit. Scale bars are 5 $\mu m$. (c) A schematic of the orbiting tubes (not to scale). The red arrow indicate the axis about which the tubes orbit.
}
\label{fig:rotating-pair}
\end{figure}

\begin{figure}[h]
\centering
\includegraphics[width=0.9\linewidth]{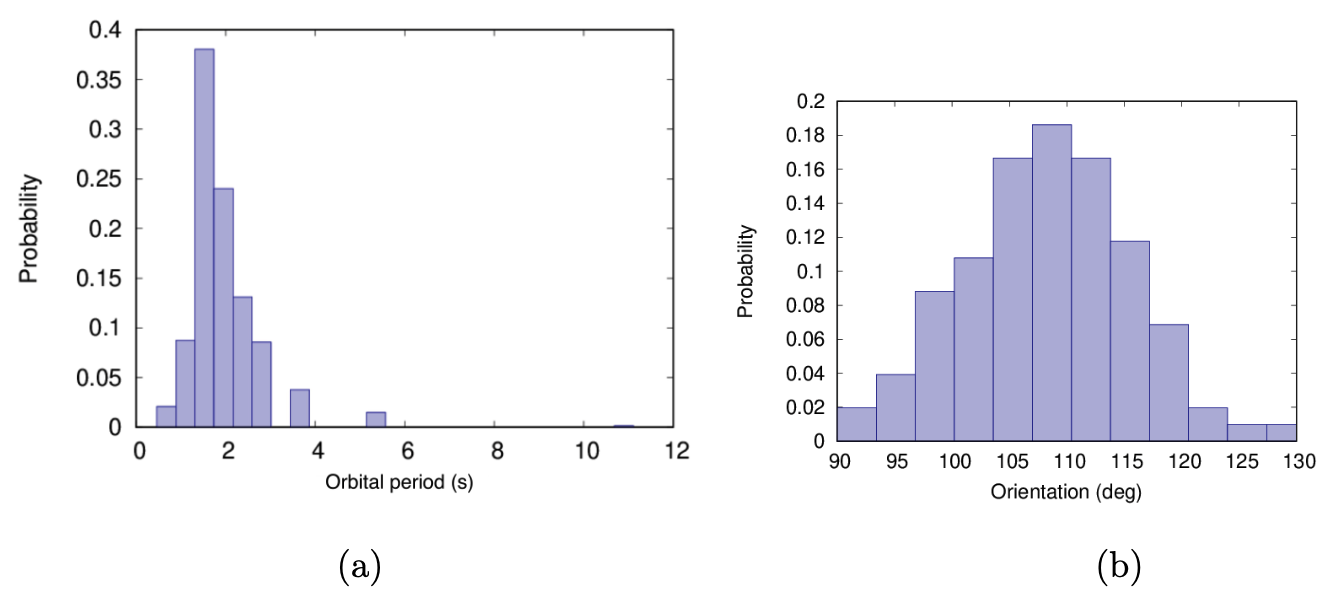}
\caption{
(a) Probability distribution of orbital frequencies of bound pairs of membrane tubes. A total of pairs were measured. (b) Probability distribution of orientation of tubes within bound pairs, measures as the obtuse angle made by each tube with respect to the equatorial tangent to the GUV.
}
\label{fig:freq-orientation}
\end{figure}

A remarkable dynamics is observed when two membrane protrusions come near each other. In such cases, the tubes form a bound pair and continuously orbit each other (see Fig. \ref{fig:rotating-pair}). Orbiting tubes appear to be tilted away from the normal to the vesicle envelope, in the direction of its motion. The tubes have been seen to remain bound for observation times of up to a minute or more. The distribution of orbital periods for such pairs is shown in Fig. \ref{fig:freq-orientation}a and that for the tilt angles is shown in Fig. \ref{fig:freq-orientation}b. Orbiting pairs are seen to remain bounded while the central axis drift around along the envelope of the vesicle (see Video-5 in Supplementary Material).  
The handedness of the orbit can be determined by observing pairs with their axis of orbit oriented normal to the plane of focus. This is particularly convenient when observing tubes near the bottom surface of the vesicle. Both senses of rotation could be observed for bound pairs. 

\section*{Discussion}

The experiments mentioned in the previous sections can be summarized as follows. Bacteria interact with the GUV membrane to create invaginations that trap them. This results in long-lived membrane tubes that protrude into the vesicle and the tubes ``diffuse'' around the surface of the vesicle. When two tubes come close to each other, they form bound pairs that orbit each other.  

\textit{E. coli} swim by spinning their flagella using rotary motors that are located at the base of each flagellum \cite{kumar2010physics, sowa2008bacterial}. When the motors spin in the counter-clockwise direction, as viewed from outside (from the tip from the flagellum towards the cell body), all the flagella form a coherent helical bundle, which spins to propel the cell body. The cell body itself will rotate with the opposite sense. Even for a given number of flgella, the drag force may vary as the cell body has a broad size distribution \cite{chattopadhyay2006swimming}. When the motors spin in the clockwise direction, the flagella becomes incoherent, splays out, and the bacterium performs a tumbling motion. The genetically modified strain we use (RP5232) are the predominant swimmers, which rarely tumble. The force generated by a swimming \textit{E. coli} is about a pico-Newton and several parameters like force, torque, rotational frequency, and efficiency have been estimated \cite{sowa2008bacterial, chattopadhyay2006swimming, armstrong2020swimming}. 

It is known that a point force applied to a lipid bilayer membrane can result in the formation of a cylindrical membrane tube \cite{derenyi2002formation}. With a membrane bending modulus $\kappa$ and in-plane tension $\sigma$, the force needed to maintain a tube can be estimated as $f = 2\pi\sqrt{2\kappa\sigma}$, and its equilibrium tube diameter as $d = \sqrt{2\kappa/\gamma}$ \cite{derenyi2002formation}. In our experiments, we observe that only floppy vesicles (low $\sigma$) form bacteria induced tubes and tubes disappear when vesicles become tense. Therefore, in low tension vesicles, tube formation can occur easily by the swimming force of bacteria, which is about 1 pN, acting against the membrane. The length of the tube depends on the available excess area, and may vary depending on how many tubes are formed.

The lengths of the tubes we observe fall within the range of 4--8 $\mu m$ which is significantly longer than the length of of the cell body of a bacterium (length ranges from 2--4 $\mu m$ and diameter is about 1 $\mu m$) \cite{chattopadhyay2006swimming}. This suggests that both the cell body and the flagella are within the tubes we observe, at least in most of the cases. The membrane invaginations show varying morphologies. When the tubes are on the shorter side of the distribution (a few microns), the tube broadens towards the base and has a wide neck (Fig. \ref{fig:single-tube}b). However, longer tubes show a narrowing of their necks (Fig. \ref{fig:single-tube}b). This narrowing is also observed near the midsections of very long tubes that carry two bacterium (Fig. \ref{fig:multiple-bacteria}). This indicates that the diameter of the thicker part of the tube is restricted by the steric hindrance caused by the bacterium. 

One of the most interesting observations is that a pair of tubes can come together and form a bound pair that orbit each other. This implies a short-range repulsion and a long-range attraction between tubes.  The observation of lipid/fluid flow near the base of the tube (Fig. \ref{fig:rotating particle}) suggests that there is a vorticity at the base of the tube, possibly generated by the rotation of the bacterial flagella. This may lead to an interaction between the two tubes which results in the orbital motion.  An additional curvature-mediated interaction can also play a role. Although elastic forces are known to cause the repulsion of inclusions in a membrane, experiments and simulations have shown that in a vesicle, with finite curvature, the interaction between inclusions can be repulsive or attractive depending on whether the inclusions are adsorbed inside or outside the vesicle \cite{Wel2016, Bahrami2018}. Simulation and experimental studies have also been performed on the adsorption of passive spherocylinders on vesicles and the resulting shape changes~\cite{laradji2023, vutukuri2024}. However, bacteria are not passive objects and, in the case reported here, they do not adhere to the membrane. The tube is most certainly the result of a pushing force exerted by a bacterium on the membrane. It'll be interesting to study the effects of spherocylinders that swim along their symmetry axis, with or without associated torque generation. It is known that the tethers pulled from a membrane with external forces results in barrier free attraction with a force proportional to the product of the forces and inversely proportional to the distance between the tethers \cite{derenyi2002formation}. It is possible that this force compounded by the repulsion between vortical flows induced by the bacteria is responsible for the bound pair formation. A detailed study to measure the various forces responsible for the stable bound pairs is underway.

Our investigation highlight the importance of physical interactions--forces, torques, and hydrodynamics--in bacteria-membrane interaction. The observation of orbiting pairs of membrane tubes demonstrates the non-trivial effects of active forces and torques in membrane mediated interactions between active chiral inclusions. Further experiments and theoretical studies are required to elucidate these mechanisms, which can be very relevant to understanding how pathogens invade host cells. 

\section*{Conflict of interest}
There are no conflicts of interest to declare 

\section*{Acknowledgements}

The authors acknowledge Hareesh Kumar and Divyang Trivedi for their help in conducting the experiments. We thank V A Raghunathan and K Vijay Kumar for discussions and the former also for help with designing experiments.


\begin{thebibliography}{23}
\providecommand{\natexlab}[1]{#1}
\providecommand{\url}[1]{\texttt{#1}}
\expandafter\ifx\csname urlstyle\endcsname\relax
  \providecommand{\doi}[1]{doi: #1}\else
  \providecommand{\doi}{doi: \begingroup \urlstyle{rm}\Url}\fi

\bibitem[Sharma et~al.(2021)Sharma, Azar, Schroder, Marques, and
  Stocco]{sharma2021active}
Vaibhav Sharma, Elise Azar, Andre~P Schroder, Carlos~M Marques, and Antonio
  Stocco.
\newblock Active colloids orbiting giant vesicles.
\newblock \emph{Soft Matter}, 17\penalty0 (16):\penalty0 4275--4281, 2021.

\bibitem[Vincent et~al.(2025)Vincent, Sreekumari, Gopalakrishnan, Vasisht, and
  Sarangi]{vincent2025curvature}
Olivia Vincent, Aparna Sreekumari, Manoj Gopalakrishnan, Vishwas~V Vasisht, and
  Bibhu~Ranjan Sarangi.
\newblock Curvature-dependent dynamics of a bacterium confined in a giant
  unilamellar vesicle.
\newblock \emph{Physical Review E}, 112\penalty0 (2):\penalty0 024408, 2025.

\bibitem[Vutukuri et~al.(2020)Vutukuri, Hoore, Abaurrea-Velasco, van Buren,
  Dutto, Auth, Fedosov, Gompper, and Vermant]{vutukuri2020active}
Hanumantha~Rao Vutukuri, Masoud Hoore, Clara Abaurrea-Velasco, Lennard van
  Buren, Alessandro Dutto, Thorsten Auth, Dmitry~A Fedosov, Gerhard Gompper,
  and Jan Vermant.
\newblock Active particles induce large shape deformations in giant lipid
  vesicles.
\newblock \emph{Nature}, 586\penalty0 (7827):\penalty0 52--56, 2020.

\bibitem[Takatori and Sahu(2020)]{takatori2020active}
Sho~C Takatori and Amaresh Sahu.
\newblock Active contact forces drive nonequilibrium fluctuations in membrane
  vesicles.
\newblock \emph{Physical review letters}, 124\penalty0 (15):\penalty0 158102,
  2020.

\bibitem[Le~Nagard et~al.(2022)Le~Nagard, Brown, Dawson, Martinez, Poon, and
  Staykova]{le2022encapsulated}
Lucas Le~Nagard, Aidan~T Brown, Angela Dawson, Vincent~A Martinez, Wilson~CK
  Poon, and Margarita Staykova.
\newblock Encapsulated bacteria deform lipid vesicles into flagellated
  swimmers.
\newblock \emph{Proceedings of the National Academy of Sciences}, 119\penalty0
  (34):\penalty0 e2206096119, 2022.

\bibitem[Nikola et~al.(2016)Nikola, Solon, Kafri, Kardar, Tailleur, and
  Voituriez]{nikola2016active}
Nikolai Nikola, Alexandre~P Solon, Yariv Kafri, Mehran Kardar, Julien Tailleur,
  and Rapha{\"e}l Voituriez.
\newblock Active particles with soft and curved walls: Equation of state,
  ratchets, and instabilities.
\newblock \emph{Physical review letters}, 117\penalty0 (9):\penalty0 098001,
  2016.

\bibitem[Robbins et~al.(1999)Robbins, Barth, Marquis, de~Hostos, Nelson, and
  Theriot]{robbins1999listeria}
Jennifer~R Robbins, Angela~I Barth, H{\'e}l{\`e}ne Marquis, Eugenio~L
  de~Hostos, W~James Nelson, and Julie~A Theriot.
\newblock Listeria monocytogenes exploits normal host cell processes to spread
  from cell to cell.
\newblock \emph{Journal of Cell Biology}, 146\penalty0 (6):\penalty0
  1333--1350, 1999.

\bibitem[Dowd et~al.(2020)Dowd, Mortuza, Bhalla, Van~Ngo, Li, Rigano, and
  Ireton]{dowd2020listeria}
Georgina~C Dowd, Roman Mortuza, Manmeet Bhalla, Hoan Van~Ngo, Yang Li,
  Luciano~A Rigano, and Keith Ireton.
\newblock Listeria monocytogenes exploits host exocytosis to promote
  cell-to-cell spread.
\newblock \emph{Proceedings of the National Academy of Sciences}, 117\penalty0
  (7):\penalty0 3789--3796, 2020.

\bibitem[Colonne et~al.(2016)Colonne, Winchell, and Voth]{colonne2016hijacking}
Punsiri~M Colonne, Caylin~G Winchell, and Daniel~E Voth.
\newblock Hijacking host cell highways: manipulation of the host actin
  cytoskeleton by obligate intracellular bacterial pathogens.
\newblock \emph{Frontiers in cellular and infection microbiology}, 6:\penalty0
  107, 2016.

\bibitem[Martinez et~al.(2018)Martinez, Siadous, and Bonazzi]{martinez2018tiny}
Eric Martinez, Fernande~Ayenoue Siadous, and Matteo Bonazzi.
\newblock Tiny architects: biogenesis of intracellular replicative niches by
  bacterial pathogens.
\newblock \emph{FEMS microbiology reviews}, 42\penalty0 (4):\penalty0 425--447,
  2018.

\bibitem[Liss and Hensel(2015)]{liss2015take}
Viktoria Liss and Michael Hensel.
\newblock Take the tube: remodelling of the endosomal system by intracellular s
  almonella enterica.
\newblock \emph{Cellular Microbiology}, 17\penalty0 (5):\penalty0 639--647,
  2015.

\bibitem[Barak and Eisenbach(1999)]{barak1999chemotactic}
Rina Barak and Michael Eisenbach.
\newblock Chemotactic-like response of escherichia coli cells lacking the known
  chemotaxis machinery but containing overexpressed chey.
\newblock \emph{Molecular Microbiology}, 31\penalty0 (4):\penalty0 1125--1137,
  1999.

\bibitem[Boban et~al.(2021)Boban, Marde{\v{s}}i{\'c}, Subczynski, and
  Raguz]{boban2021giant}
Zvonimir Boban, Ivan Marde{\v{s}}i{\'c}, Witold~Karol Subczynski, and Marija
  Raguz.
\newblock Giant unilamellar vesicle electroformation: What to use, what to
  avoid, and how to quantify the results.
\newblock \emph{Membranes}, 11\penalty0 (11):\penalty0 860, 2021.

\bibitem[Dai et~al.(2023)Dai, Tang, Zhang, Li, He, Han, and Wang]{dai2022lipid}
Shaoying Dai, Xiaoyu Tang, Na~Zhang, Haofei Li, Chengzhi He, Yuchun Han, and
  Yilin Wang.
\newblock Lipid giant vesicles engulf living bacteria triggered by minor
  enhancement in membrane fluidity.
\newblock \emph{Nano letters}, 23\penalty0 (1):\penalty0 371--379, 2023.

\bibitem[Kumar and Philominathan(2010)]{kumar2010physics}
M~Siva Kumar and P~Philominathan.
\newblock The physics of flagellar motion of e. coli during chemotaxis.
\newblock \emph{Biophysical reviews}, 2\penalty0 (1):\penalty0 13--20, 2010.

\bibitem[Sowa and Berry(2008)]{sowa2008bacterial}
Yoshiyuki Sowa and Richard~M Berry.
\newblock Bacterial flagellar motor.
\newblock \emph{Quarterly reviews of biophysics}, 41\penalty0 (2):\penalty0
  103--132, 2008.

\bibitem[Chattopadhyay et~al.(2006)Chattopadhyay, Moldovan, Yeung, and
  Wu]{chattopadhyay2006swimming}
Suddhashil Chattopadhyay, Radu Moldovan, Chuck Yeung, and XL~Wu.
\newblock Swimming efficiency of bacterium escherichia coli.
\newblock \emph{Proceedings of the National Academy of Sciences}, 103\penalty0
  (37):\penalty0 13712--13717, 2006.

\bibitem[Armstrong et~al.(2020)Armstrong, Nieminen, Stilgoe, Kashchuk, Lenton,
  and Rubinsztein-Dunlop]{armstrong2020swimming}
Declan~J Armstrong, Timo~A Nieminen, Alexander~B Stilgoe, Anatolii~V Kashchuk,
  Isaac~CD Lenton, and Halina Rubinsztein-Dunlop.
\newblock Swimming force and behavior of optically trapped micro-organisms.
\newblock \emph{Optica}, 7\penalty0 (8):\penalty0 989--994, 2020.

\bibitem[Der{\'e}nyi et~al.(2002)Der{\'e}nyi, J{\"u}licher, and
  Prost]{derenyi2002formation}
Imre Der{\'e}nyi, Frank J{\"u}licher, and Jacques Prost.
\newblock Formation and interaction of membrane tubes.
\newblock \emph{Physical review letters}, 88\penalty0 (23):\penalty0 238101,
  2002.

\bibitem[Van Der~Wel et~al.(2016)Van Der~Wel, Vahid, {\v{S}}ari{\'c}, Idema,
  Heinrich, and Kraft]{Wel2016}
Casper Van Der~Wel, Afshin Vahid, An{\dj}ela {\v{S}}ari{\'c}, Timon Idema,
  Doris Heinrich, and Daniela~J Kraft.
\newblock Lipid membrane-mediated attraction between curvature inducing
  objects.
\newblock \emph{Scientific reports}, 6\penalty0 (1):\penalty0 32825, 2016.

\bibitem[Bahrami and Weikl(2018)]{Bahrami2018}
A~H Bahrami and T.~R. Weikl.
\newblock Curvature-mediated assembly of janus nanoparticles on membrane
  vesicles.
\newblock \emph{Nano letter}, 18:\penalty0 1259--1263, 2018.

\bibitem[Sharma et~al.(2023)Sharma, Zhu, Spangler, Carrillo, and
  Laradji]{laradji2023}
Abash Sharma, Yu~Zhu, Eric~J Spangler, Jan-Michael~Y Carrillo, and Mohamed
  Laradji.
\newblock Membrane-mediated dimerization of spherocylindrical nanoparticles.
\newblock \emph{Soft Matter}, 19\penalty0 (8):\penalty0 1499--1512, 2023.

\bibitem[Van~der Ham et~al.(2024)Van~der Ham, Agudo-Canalejo, and
  Vutukuri]{vutukuri2024}
Stijn Van~der Ham, Jaime Agudo-Canalejo, and Hanumantha~Rao Vutukuri.
\newblock Role of shape in particle-lipid membrane interactions: from surfing
  to full engulfment.
\newblock \emph{ACS nano}, 18\penalty0 (15):\penalty0 10407--10416, 2024.

\end{thebibliography}
	
\end{document}